\documentclass[english,aps]{revtex4}
\usepackage[T1]{fontenc}
\usepackage[latin9]{inputenc}
\usepackage[letterpaper]{geometry}
\usepackage{color}
\usepackage{babel}
\usepackage{array}
\usepackage{float}
\usepackage{booktabs}
\usepackage{multirow}
\usepackage[unicode=true,pdfusetitle,
 bookmarks=true,bookmarksnumbered=false,bookmarksopen=false,
 breaklinks=false,pdfborder={0 0 1},backref=false,colorlinks=true]
 {hyperref}
\hypersetup{
 urlcolor=blue}
\usepackage{breakurl}

\makeatletter

\providecommand{\tabularnewline}{\\}

\@ifundefined{textcolor}{}
{%
 \definecolor{BLACK}{gray}{0}
 \definecolor{WHITE}{gray}{1}
 \definecolor{RED}{rgb}{1,0,0}
 \definecolor{GREEN}{rgb}{0,1,0}
 \definecolor{BLUE}{rgb}{0,0,1}
 \definecolor{CYAN}{cmyk}{1,0,0,0}
 \definecolor{MAGENTA}{cmyk}{0,1,0,0}
 \definecolor{YELLOW}{cmyk}{0,0,1,0}
 }

\makeatother

\begin{document}

\title{Gender and Sexual Diversity Issues in Physics: The Audience Speaks}

\author{N. Ackerman}

\affiliation{Department of Physics, Stanford University, Stanford, California.
94305}

\author{T. J. Atherton}

\email{timothy.atherton@tufts.edu}

\selectlanguage{english}%

\affiliation{Department of Physics and Astronomy, Tufts University, 4 Colby Street,
Medford, Massachusetts. 02144}

\author{W. Deconinck}

\affiliation{Department of Physics, College of William \& Mary, Williamsburg,
Virginia. 23185}

\author{M. L. Falk}

\affiliation{Department of Materials Science and Engineering, Department of Mechanical
Engineering and Department of Physics and Astronomy, Johns Hopkins
University, Baltimore, Maryland. 21218}

\author{S. Garmon}

\affiliation{Chemical Physics Theory Group, Department of Chemistry, University
of Toronto, 80 St. George Street, Toronto, Ontario, Canada. M5S 3H6}

\author{E. Henry}

\affiliation{Department of Physics, University of California, Berkeley, California.
94720}

\author{E. Long}

\affiliation{Department of Physics, Kent State University, Kent, Ohio. 44242}

\maketitle

\section{Introduction}

The session \href{http://meetings.aps.org/Meeting/MAR12/SessionIndex3/?SessionEventID=168349}{``Gender and Sexual Diversity Issues in Physics''}
took place at the APS March Meeting 2012 in Boston, MA on the 28th
of February\citep{session}. The session consisted of four invited
talks from invited speakers Susan Rankin (in fact presented by Eric
Patridge of \href{http://www.ostem.org/}{oSTEM} and Ram\' on Barthelemy),
Michael Ramsey-Musolf, Janice Hicks and Elena Long, followed by a
panel discussion where the invited speakers were joined by Ted Hodapp.
Over 120 individuals attended some part of the session according to
a rough headcount by the organizers and perhaps half of those stayed
for the whole time. 

Since a theme of the session was on gathering data and creating conversation,
a survey (included as Appendix A) was designed by Wouter Deconinck
and Savannah Garmon with input from Tim Atherton, Elena Long, Ned
Henry, Nicole Ackerman and Michael Falk. The questions were intended
to be inclusive, i.e. to invite as broad a range of thoughts on the
session as possible, and also to allow participants to give a preliminary
indication of their thoughts on the role of the APS. A total of 43
responses were received. Above all, it appears that the session was
viewed very positively, being characterized by respondents as {}``\emph{Wonderful!'',
{}``Important'', {}``Informative'' }and\emph{ {}``Strengthening''.}

The purpose of the present paper is to document some of the ideas
presented in the session in section II and to report the responses
from the audience in the survey in sections III\textemdash{}VI. To
assemble a coherent narrative from the survey, the questions and responses
have been divided into four broad areas: demographic data, thoughts
on the session itself, general thoughts on steps toward progress and,
finally, action items. In these sections, only light interpretation
has been added as we believe many of the comments speak for themselves.
All responses have been included in this report except for comments
that were nearly identical in wording; it is noted where such responses
have been omitted to avoid repetition. Not all questions were answered
by all respondents and so tallies of replies do not typically add
up to the total number of responses. The report concludes in section
VII by drawing a series of recommendations from the data.

\section{The Session}

The session began with a talk entitled\emph{ \href{http://meetings.aps.org/link/BAPS.2012.MAR.J20.1}{``The State of Higher Education for STEM LGBTQQ Faculty/Staff,''}}
prepared by Susan Rankin\citep{rankin} and presented by Eric Patridge
and Ram\' on Barthelemy. The talk presented results from a national
climate survey written by Prof. Rankin and conducted by the \href{http://www.campuspride.org/research/qrihe.htm}{Q Research Institute for Higher Education (QRIHE)}
and \href{http://www.campuspride.org/}{Campus Pride}\citep{qrhe,campuspride}.
The survey focused on observation and experience of exclusionary behavior
and the effects of this behavior on faculty and staff retention. According
to the survey results, LGBT%
\footnote{We use the acronym LGBT (Lesbian Gay Bisexual Transgender) here in
this report as it is widely recognized; we do not mean to exclude
any other identities. Other longer but more inclusive versions of
this acronym also exist, such as LGBTIQQAP+ (Lesbian Gay Bisexual
Transgender Intersex Queer Questioning Allies Pansexual and all others). %
} faculty members in STEM fields are more likely to be out than their
colleagues in other fields. Approximately half of all STEM faculty
participating in the survey reported observing exclusionary behavior,
while 20\% report had experienced it. Of the STEM faculty, 50\% had
considered leaving their departments due to this behavior and 70\%
report that the exclusionary behavior came from the administration
rather than the student body. The faculty members who reported experiencing
exclusionary behavior were far more likely to consider leaving their
institution or the field. The presentation highlighted the necessity
for more thorough research into the climate in STEM fields and the
practical necessity for employers in STEM fields to create an environment
in which LGBT faculty feel welcome in order to increase retention
rates. 

Michael Ramsey-Musolf presented the second talk of the session entitled
\emph{\href{http://meetings.aps.org/link/BAPS.2012.MAR.J20.2}{``Shattering the Lavender Ceiling: Sexual Minorities in Physics''}\citep{ramsey}.}
Prof. Ramsey-Musolf emphasized the role that scientists have historically
played in human rights struggles, noting as an example the work of
the APS \href{http://www.aps.org/about/governance/committees/cifs/index.cfm}{Committee on International Freedom of Scientists}\citep{apscifs}.
He stated explicitly his own perception of problems faced by the LGBT
community in STEM fields: a hostile climate, exclusion from opportunities
and non-receptivity to concerns by those in positions of relative
power. He observed that these problems are caused by invisibility,
ignorance and an absence of data and they result in isolation, alienation,
loss of talent from field and perpetuation of stereotypes. Ramsey-Musolf
expressed optimism that scientists as a group want to do the right
thing and called on the community to share their stories, adopt non-discrimination
policies openly, establish a task force explicitly dedicated to LGBT
rights and to create more resources such as {}``out lists'' to help
raise the bar for the climate in our fields.

Professor Janice Hicks shared her personal story as a physical chemist
and a lesbian in her talk\emph{ \href{http://meetings.aps.org/link/BAPS.2012.MAR.J20.3}{``Why Awareness of LGBT issues in the Physics Community Makes Sense''}}\citep{hicks}.
Prof. Hicks was not out as a lesbian to her colleagues until she left
her faculty position at age forty, having been uncomfortable with
the climate throughout her career. Official acceptance of her identity
at her new position at the NSF was an extremely important factor in
her decision to stay there. Moving forward, Hicks emphasized the need
for education to combat stereotypes, for state legal protections regarding
employment discrimination and for climate surveys that hold individual
institutions accountable and provide public relations incentives to
make students and employees feel welcome. 

The final talk was given by Elena Long, entitled \emph{\href{http://meetings.aps.org/link/BAPS.2012.MAR.J20.4}{``Physics Climate as Experienced by LGBT+ Physicists''}\citep{long}.}
This talk discussed the history of the organizing committee for this
session and the website \href{http://lgbtphysicists.x10hosting.com/}{LGBTIQQAP+ Physicists}
that ze set up\citep{lgbtphysicists}. Long also presented the results
of a climate survey conducted through the Forum of Graduate Student
Affairs\citep{longcitation}. The survey focused on reasons why students
in STEM fields feel unsafe or uncomfortable. Nearly all of the survey
respondents who identify outside of the gender binary reported feeling
unsafe in their workplace for one reason or another. Furthermore,
racial and religious minorities, especially people with disabilities,
were more likely to report feeling unsafe due to gender related issues,
highlighting the intersectionality of social justice issues in our
fields. Long concluded by emphasizing the need for more data, more
education and more awareness in our fields. 

A panel discussion concluded the session, focusing principally on
how best to address the issues raised in the preceding talks. The
group discussed the challenges of presenting problems which are often
qualitative by nature to a community dedicated to quantitative research.
Funding sources for further research and organization were discussed,
with an emphasis on securing government recognition as an \textquotedblleft{}underrepresented
minority\textquotedblright{} and on increasing awareness among philanthropic
individuals and organizations which would be otherwise unaware of
this issue. Michael Ramsey-Musolf advocated for creating public \textquotedblleft{}out
lists\textquotedblright{} to decrease the isolation of LGBT identified
people in STEM fields, pointing out that fixing the problem is both
a practical necessity for the field and a basic human rights issue.
The possibility of organizing mentorship programs focused on sexual
and gender identities through APS was raised. Ted Hodapp of the APS
encouraged members of the APS to help improve the climate for LGBT
physicists noting that, as a large and primarily volunteer organization,
the APS relies on its members to advance such causes. The panel ended
with a personal call for everyone in attendance to bring their full
intellectual power to bear on this problem, to hold organizations
and powerful individuals personally accountable for their failures
and to confront prejudice and discrimination directly and personally.

\section{Demographic}

We now turn to our analysis of the results of the survey. Only half
of the respondents identified as a sexual or gender minority (table
\ref{tab:q1}). The single {}``other'' respondent explicitly identified
as {}``F'' (for female). Those who considered themselves as belonging
to a minority were invited to explain how they identified: the identities
provided are listed in table \ref{tab:q2}; note that the responses
are not mutually exclusive and some respondents provided more than
one of these identities. Two respondents chose to explicitly add {}``male''
having identified as {}``gay''. Additionally, three respondents
who did not identify as belonging to a sexual/gender minority chose
to identify themselves as an {}``ally''.

\begin{table}
\centering{}\caption{\label{tab:q1}Do you identify as a sexual/gender minority?}
\begin{tabular}{cc}
\hline 
Yes & 21\tabularnewline
No & 20\tabularnewline
Other & 1\tabularnewline
No response & 1\tabularnewline
\hline 
\end{tabular}
\end{table}
\begin{table}
\centering{}\caption{\label{tab:q2}How do you identify?}
\begin{tabular}{cc}
\hline 
Gay & 10\tabularnewline
Lesbian & 2\tabularnewline
Female%
\footnote{We note there is much ambiguity in this identity. One such respondent
wrote {}``I identify as a gender minority being a female physicist''.%
} & 2\tabularnewline
Trans & 3\tabularnewline
Bi & 2\tabularnewline
Asexual & 1\tabularnewline
Inter & 1\tabularnewline
Queer & 3\tabularnewline
\hline 
\end{tabular}
\end{table}

The majority of participants were students, as is evident from the
question on career stage displayed in table \ref{tab:q3}. Two students
explicitly noted that they were graduate students; one as an undergraduate.
Of those who selected {}``other'', three identified themselves as
press or science writers and one person as a researcher. 

\begin{table}
\centering{}\caption{\label{tab:q3}At what stage of your career are you?}
\begin{tabular}{cc}
\hline 
Student & 25\tabularnewline
Postdoc & 4\tabularnewline
Pre-tenured faculty & 1\tabularnewline
Tenured faculty & 7\tabularnewline
Other & 5\tabularnewline
\hline 
\end{tabular}
\end{table}

\section{The Session\label{sec:The-Session}}

To assist in understanding how people came to know of the session,
a question was posed to this effect and is reported in table \ref{tab:q9}.
While it is clear that advertising at the meeting is crucial, several
people specifically mentioned the Epitome\citep{epitome}. The Google
group LGBT+ Physicists\citep{google} was also mentioned. 

\begin{table}
\centering{}\caption{\label{tab:q9}When and how did you hear about the panel?}
\begin{tabular}{cc}
\hline 
At the meeting & 29\tabularnewline
From a friend & 3\tabularnewline
Specific announcement & 3\tabularnewline
Other & 7\tabularnewline
\hline 
\end{tabular}
\end{table}

The session was clearly very much appreciated by the audience (see
table \ref{tab:q10}), although a selection bias in those who filled
out the survey should be considered. 

\begin{table}
\centering{}\caption{\label{tab:q10}Do you think it was worth attending the session?}
\begin{tabular}{cc}
\hline 
Yes & 33\tabularnewline
{}``Sort of''%
\footnote{In keeping with the desire to allow respondents to choose their own
words, no prescribed response was provided (i.e. Yes/No); the respondent
is therefore quoted verbatim.%
} & 1\tabularnewline
\hline 
\end{tabular}
\end{table}

Respondents were then asked to describe the session, which elicited
universally positive responses%
\footnote{One respondent wished that more effort had been made to reach out
to allies; this reply is discussed at the end of section V in the
context of a similar reply to another question.%
} like
\begin{quotation}
\emph{{}``Enlightening''}

\emph{{}``Important''}

\emph{{}``Nice balance''}

\emph{{}``informative \& strengthening''}

\emph{{}``attentive/sensitive/nuanced commentary healthy discussion''}

\emph{{}``Interesting''}

\emph{{}``Wonderful! Good balance of information/data and personal
stories.''}

\emph{{}``very useful. Great information.''}

\emph{{}``It has brought out assumptions, issues from the background
into the open.''}

\emph{{}``worthwhile, especially in terms of networking''}
\end{quotation}
What were they taking away? 
\begin{itemize}
\item \textbf{Need for data:} \emph{{}``The need for much better numbers
on the size and needs of the community'' {[}many nearly identical
comments{]}}
\item \textbf{Role models:} {}``\emph{There are other people like me who
are further in their careers. I know a number of queer students or
allies that are students, but I have no visible, living queer scientist
role models.''}
\item \textbf{Awareness:} \emph{{}``I learned more about specific issues
that I hadn't thought about before, esp. for trans people''}\\
\emph{}\\
\emph{{}``Paying more attention to the manifestations of exclusive
behavior and the lavender ceiling, mentioning/welcoming sexual minorities
with students.''}\\
\emph{}\\
\emph{{}``More ideas about how to create blanket verbiage and approaches
that address discriminatory policies and practices applied to all
minority, underrepresented members (ethnic, gender, sexual). I'd like
to take these ideas to my home institution.''}
\item \textbf{Importance of allies:} \emph{{}``need more info for non-LGBTQ
on how to be more involved''}
\item \textbf{Pride:} \emph{{}``More specifically a `Physics Pride'''}
\item \textbf{{}``We need to continue the discussion''}
\end{itemize}
Respondents were also asked for things that they agreed or disagreed
with and for any other other comments about any of the talks. None
of these questions elicited many responses. One respondent felt that
\begin{quotation}
\emph{{}``Having an `out list' is, I think, dangerous.''}
\end{quotation}
Several people mentioned that they were \emph{{}``deeply} \emph{moved''}
by Prof. Ramsey-Musolf's brief account of Alan Turing's story and
the subsequent apology by Gordon Brown, former Prime Minister of the
UK. Several others mentioned that speaking openly and hearing stories
was beneficial. Some of the responses touched on themes addressed
earlier, e.g.
\begin{quotation}
\emph{{}``the idea of discussion/teaching of history \& philosophy
of science; in general, having courses considering the community of
scientists/science could be helpful.''}
\end{quotation}
One respondent felt that
\begin{quotation}
\emph{{}``I think the name of this session isn't quite right. It
is unclear whether it includes women in science issues since women
are a `gender minority' in physics. Something that specifically said
LGBT+ or something like that might be better or just saying sexuality
and gender instead. It's a more well known phrase.\textquotedbl{}}
\end{quotation}
The value of the session was eloquently summarized by another respondent:
\begin{quotation}
\emph{{}``I think showing successful academics who are out is an
important way to dispel the notion that being out precludes or limits
one's professional opportunities. So thank you to your speakers.''}
\end{quotation}

\section{Moving Forward\label{sec:Moving-Forward}}

An indication of the \emph{status quo} may be seen in table \ref{tab:q4},
where it is apparent that most respondents seldom or never discuss
sexual or gender issues with other scientists. Four respondents qualified
their responses to indicate a limitation on the circle of people they
did discuss these issues with, specifying friends or other female
physicists. 

\begin{table}
\centering{}\caption{\label{tab:q4}How often do you discuss sexual and gender issues with
other scientists?}
\begin{tabular}{cc}
\hline 
Daily & 0\tabularnewline
Weekly & 6\tabularnewline
Monthly & 5\tabularnewline
Seldom & 25\tabularnewline
Never & 7\tabularnewline
\hline 
\end{tabular}
\end{table}

Perhaps because of this, there was almost unanimous support for increased
visibility of sexual and gender minorities (see table \ref{tab:q5}).
Two respondents added that this was also desirable internationally
(this was also raised in the question afterwards on how to achieve
the visibility). One dissenting respondent qualified their response
thus
\begin{quotation}
\emph{{}``I do not agree that the visibility is the primary goal
of the LGBT+ community. I think the primary goal is to make people
really don't care about }any\emph{ factors except for professionalism
of a physicist.''}
\end{quotation}
\begin{table}
\centering{}\caption{\label{tab:q5}Do you consider it to be an important goal for sexual
and gender minorities to take on a more visible role in science nationally?}
\begin{tabular}{cc}
\hline 
Yes & 38\tabularnewline
No & 2\tabularnewline
Uncertain & 1\tabularnewline
\hline 
\end{tabular}
\end{table}

Participants were asked if they had ideas about how to increase the
visibility. Quite a few ideas were presented, often by more than one
person. Some themes grouped together with representative comments:
\begin{itemize}
\item \textbf{Being out}: \emph{{}``It shouldn't be their responsibility
but they can help dispel stereotypes about the minorities and of science{[}...{]}
Being out (and vocal) might help accomplish it.''}
\item \textbf{Education (within physics):} \emph{{}``Design strategies
to improve the education of students, mainly undergraduate, on these
issues''}
\item \textbf{Education/publicity (of the general public):} \emph{{}``I
think {[}...{]} public service messages highlighting queer scientists
and discussing the issue would be good, as well as putting queer characters
in children's programs.''}\\
\emph{}\\
\emph{{}``perhaps reaching out LGBT+ serving media outlets with suggested
contributed content highlighting LGBT+ serving scientists \& their
work, or doing the same with local news outlets.''}
\item \textbf{Leaders:} \emph{{}``I think making more of an effort to represent
open sexual minorities to represent leadership roles.''}\\
\emph{}\\
\emph{{}``Get them in more leadership positions. More visibility!''}
\item \textbf{Mentoring: }\emph{{}``More active mentoring by faculty and
older graduate students. Most scientists are accepting and liberal
and many are allies, but it's often unclear who they are.''}
\item \textbf{Grassroots effort: }\emph{{}``While I do want acceptance
and integration, it is hard to imagine this acceptance and integration
to occur primarily through centralized efforts. From my own experience,
individually speaking with colleagues and being locally visible goes
a lot further. Perhaps supporting the individuals efforts }is\emph{
a good role for a central organizing group.''}
\item \textbf{The session itself:}\emph{ {}``This discussion is a good
start.''}
\end{itemize}
Respondents were then invited to give other suggestions on how to
move the conversation forward regarding the role of sexual and gender
minorities in science. While there were fewer responses to this question,
they were generally very detailed. By theme:
\begin{itemize}
\item \textbf{Collect data: }\emph{{}``Find out the percentage of LGBT
people in the science community''}
\item \textbf{Empower allies:} \emph{{}``Start by making it an issue. Speaking
as an `ally', I have trouble knowing what I can do to help. What can
those of us who are part of the majority do? We need input on that.''}\\
\emph{}\\
\emph{{}``Improve the comfort level of non-LGBT faculty''}\\
\emph{}\\
\emph{{}``I think sexuality and gender minority issues need to be
made an issue that heteronormative people discuss.''}
\item \textbf{Share stories, data and practices:} \emph{{}``Tell more stories
about minority members at research institutions and at larger conferences
and workshops, present more statistics and data about how other (even
non-academic) organizations promote environments of equality as models
to follow and present data about how significant of a percentage minorities
comprise in populations, social groups to emphasize how `normal' minority
members are in society as a whole (by `normal' I mean not fringe or
niche or rare) and thus merit proper representation.''}\\
\emph{}\\
\emph{{}``Shared experiences and development of best practices may
help.''}
\item \textbf{Scholarships:} \emph{{}``It might be helpful to have similar
{[}to scholarship for women in physics{]} programs for LGBT etc. students
or if there are not enough LGBT people in a given community LGBT issues
and women's issues because although there are plenty of unique experiences
to each community there are also shared experiences, many around the
kind of stereotypes part of both group's issues include the prevalence
of the idea of science being a `masculine' activity.''}
\item \textbf{More opportunities to talk}: \emph{{}``NETWORK!''}\\
\emph{}\\
\emph{{}``More sessions like this! I'm amazed at how little these
issues are being discussed and how few scientists are out. I suspect
that a lot of people }think\emph{ that the academy is a liberal atmosphere
and thus - we're doing well, but that isn't enough (and often isn't
even true).''}
\end{itemize}
For future sessions, it is notable that one person included in their
reply to this question that
\begin{quotation}
\emph{{}``I felt I was outing myself by simply walking into this
room and was afraid to attend the talks.''}
\end{quotation}
and in response to another question, a similar note was struck
\begin{quotation}
\emph{{}``the assumption seemed to be that most of the audience was
LGBTQ. I think a greater effort to reach out to allies would have
been helpful.''}
\end{quotation}
One respondent left a detailed comment
\begin{quotation}
\emph{{}``I find it ironic that you are discriminatory/biased toward
academia. Are there no LGBT+ issues among physicists in industry or
national labs? Most physics students do not stay in academia anyway.''}
\end{quotation}
suggesting that future work should include exploration of this wider
focus (see recommendations). 

\begin{table}[h]
\centering{}\caption{\label{tab:q8}Rank the following actions that the APS could take
to be more inclusive of LGBT+ people in physics. Models A, B, C, D
reflect different scoring systems used to obtain the overall rank
(see Appendix B); numbers indicate priority of action, with $1$ corresponding
to highest priority.}
\begin{tabular}{>{\raggedright}p{3in}>{\centering}p{0.5cm}>{\centering}p{0.5cm}>{\centering}p{0.5cm}>{\centering}p{0.5cm}}
\toprule 
\multirow{2}{3in}{\center Action} & \multicolumn{4}{c}{Overall Rank}\tabularnewline
\cmidrule{2-5} 
 & \emph{A} & B & C & D\tabularnewline
\midrule
\everypar{\hangindent1em \hangafter1}\raggedright Sponsor a conference
to explore the status of LGBT people in physics. & 6 & 5 & 5 & 6\tabularnewline
\addlinespace
\everypar{\hangindent1em \hangafter1}\raggedright Set up a committee
along the lines of the CSWP (Committee of the status of women in physics)
to represent LGBT people in physics. & 3 & 3 & 3 & 3\tabularnewline
\addlinespace
\everypar{\hangindent1em \hangafter1}\raggedright Conduct or facilitate
a national survey to gauge the needs of and present climate for LGBT
people in physics. & 1 & 1 & 1 & 1\tabularnewline
\addlinespace
\everypar{\hangindent1em \hangafter1}\raggedright Set up a subcommittee
within the broader Committee on Minorities. & 5 & 4 & 4 & 4\tabularnewline
\addlinespace
\everypar{\hangindent1em \hangafter1}\raggedright Develop \textquotedbl{}best
practice\textquotedbl{} guidelines for Department Chairs to create
an LGBT-inclusive environment. & 2 & 2 & 2 & 2\tabularnewline
\addlinespace
\everypar{\hangindent1em \hangafter1}\raggedright Something else
(with space for ideas) & 4 & 6 & 6 & 5\tabularnewline
\bottomrule
\addlinespace
\end{tabular}
\end{table}

\section{Action\label{sec:Action}}

Respondents were invited to rank some possible actions that the APS
could take to be more inclusive of LGBT+ people in physics. Since
this is a vote, there is no perfect procedure%
\footnote{Due to Arrow's impossibility theorem%
} for ranking these preferences; we used four different methods as
described in Appendix B. The order of preference of the various options
are displayed in table \ref{tab:q8}. Notice that the ordering of
the top three preferences does \textbf{not} change under our selection
of reasonable ranking strategies. 

It is clear the strongest preference for immediate action is for a
national survey, with strong support for {}``best practice guidelines''
and also support for representation; the relative strength of the
recommendations at the conclusion of this report is intended to reflect
this ranking. Respondents seemed to prefer the idea of a committee
along the lines of CSWP to the idea of a subcommittee, although one
respondent noted \emph{{}``surely the same''} for these two options.
For other possible activities, respondents suggested:
\begin{quotation}
\emph{{}``Mentoring programs''}

\emph{{}``A group for students and others to get together \& organize.''}

\emph{{}``Publish brochures on contributions of queer/LGBT+ scientists.
Role models are important in forming positive identities.''}

\emph{{}``Make explicit statements online''}

\emph{{}``LGBT networking/social events''}

{[}referring to best practice{]} \emph{{}``Actually, this is really
important. Even just a passing comment about how the department is
a welcoming and safe space could really affect how the atmosphere
is.''}
\end{quotation}

\section{Recommendations}

By drawing together the choices made by respondents to the survey
and also issues raised in responses, we conclude that the APS, partner
organizations and the community of physicists who identify as a sexual
or gender minority or an ally should:
\begin{enumerate}
\item Design and conduct, or support, a national survey to understand the
experiences and climate for sexual and gender minorities in the physics
community. Possible partners include, for instance, the American Institute
of Physics which already carries out extensive surveys of Physicists\citep{aip}.
Respondents to the present survey endorsed this as the \textbf{highest
priority}\emph{ }of possible outcomes of the session\emph{ }at March
meeting 2012\emph{ {[}see section \ref{sec:Action}{]}. }
\item Consult with department chairs and the private sector to produce and
make available a set of {}``best practices''.\emph{ }This was endorsed
as the second highest priority by survey respondents\emph{ {[}see
section \ref{sec:Action}{]}.}
\item Explore ways to represent people who identify as belonging to a sexual
or gender minority who are not already represented by COM and CSWP;
consider how the structure of these committees might facilitate exploration
of intersectionality while continuing to address the unique issues
faced by each kind of minority. This was endorsed as the third highest
priority by survey respondents \emph{{[}see section \ref{sec:Action}{]}.}
\item Empower allies with resources and actionable items (e.g. encouraging
them to participate in their institution's version of Safe Zone).
\emph{{[}see section \ref{sec:Moving-Forward}{]}}
\item Explore the status of sexual and gender minorities in industry and
national labs and make people in these career paths feel included
in the discussion and in future events. \emph{{[}see section \ref{sec:Moving-Forward}{]}}
\item Facilitate, provide and publicize \emph{fora} for discussion of sexual
and gender minority issues in physics, both online and at future APS
conferences and other relevant opportunities. \emph{{[}see sections
\ref{sec:Moving-Forward} and \ref{sec:Action}{]}}
\item Facilitate, provide and publicize networking events for sexual and
gender minority physicists. \emph{{[}see section \ref{sec:Moving-Forward}
and \ref{sec:Action}{]}}
\item Identify partners who might facilitate similar movements in other
countries (e.g. Institute of Physics). Relate experiences and provide
resources to support these efforts. \emph{{[}see section }\ref{sec:Moving-Forward}\emph{{]}} 
\item Develop the website LGBTIQQAP+ Physicists as a central resource for
information. Identify additional channels for advertising future similar
sessions \emph{{[}see section \ref{sec:The-Session}{]}}.
\item At future events, vocalize the fact that the room includes both minority
and non-minority people and that presence is no indication of minority
status. Make the safety of the space explicit\emph{ {[}see section
\ref{sec:Moving-Forward}{]}.}\end{enumerate}
\begin{acknowledgments}
The organizing committee for the session wish to reiterate their thanks
to APS, the CSWP and the COM for their support in hosting the session
and to the speakers and attendees. We also wish to thank the Tufts
University LGBT Center for assistance in preparing this report.

\end{acknowledgments}

\section*{APPENDIX A. The Survey}
\begin{enumerate}
\item Do you identify as a sexual/gender minority (gay, lesbian, trans,
etc.)? 
\item If so, how do you identify? 
\item At what stage of your career are you? 

\begin{enumerate}
\item Student
\item Postdoc
\item Pre-tenured faculty
\item Tenured faculty
\item Other \emph{{[}Space provided{]}}
\end{enumerate}
\item How often do you discuss sexual and gender issues with other scientists?

\begin{enumerate}
\item Daily
\item Weekly
\item Monthly
\item Seldom
\item Never 
\end{enumerate}
\item Do you consider it to be an important goal for sexual and gender minorities
to take on a more visible role in science nationally?
\item If so, do you have any thoughts on how that might be accomplished?
\item Do you have any other suggestions on how to move the conversation
forward regarding the role of sexual and gender minorities in science? 
\item Rank the following actions that the APS could take to be more inclusive
of LGBT+ people in physics 1: most important, 6: least important

\begin{enumerate}
\item Sponsor a conference to explore the status of LGBT people in physics. 
\item Set up a committee along the lines of the CSWP (Committee of the status
of women in physics) to represent LGBT people in physics. 
\item Conduct or facilitate a national survey to gauge the needs of and
present climate for LGBT people in physics.
\item Set up a subcommittee within the broader Committee on Minorities. 
\item Develop \textquotedbl{}best practice\textquotedbl{} guidelines for
Department Chairs to create an LGBT-inclusive environment.
\item Something else: \emph{{[}Space provided{]}}
\end{enumerate}
\item When and how did you hear about the APS panel on sexual and gender
diversity?

\begin{enumerate}
\item At the meeting
\item From a friend, 
\item Specific announcement
\item Other: \emph{{[}Space provided{]}}
\end{enumerate}
\item Do you think it was worth attending the session?
\item How would you describe the session, in general terms?
\item What is one thing that you are taking away from this session?
\item Was there something presented that you particularly agreed with?
\item Was there something presented that you disagreed with?
\item Do you have any specific comments on a particular talk (feel free
to respond to any, all or none)?
\end{enumerate}

\section*{APPENDIX B. Ranking Preferences}

As noted in section \ref{sec:Action}, there's no clear optimal procedure
to discern a rank for candidate actions from a set of votes. Generically
one can assign scores $1^{st}\to w_{1}$, $2^{nd}\to w_{2}$, etc.
and sum the scores for each option over the set of responses. We used
for different weighting schemes $A=(1,0,0,0,0,0)$, $B=(1,\frac{1}{2},\frac{1}{4},\frac{1}{8},\frac{1}{16},0)$,
$C=(1,\frac{4}{5},\frac{3}{5},\frac{2}{5},\frac{1}{5},0)$ and $D=(\frac{1}{2},\frac{1}{2},0,0,0,0)$,
where each weighting vector was additionally normalized, to produce
an overall score for each action; these are displayed in \ref{tab:q8-1}.
The ranking deduced from these scores is displayed in table \ref{tab:q8}. 

\begin{table}[H]
\centering{}\caption{\label{tab:q8-1}Rank the following actions that the APS could take
to be more inclusive of LGBT+ people in physics. Models A, B, C, D
reflect different scoring systems for the overall rank. }
\begin{tabular}{>{\raggedright}m{3in}>{\centering}m{0.5cm}>{\centering}p{0.5cm}>{\centering}p{0.5cm}>{\centering}p{0.5cm}}
\toprule 
\multirow{2}{3in}{\center Action} & \multicolumn{4}{c}{Overall Score}\tabularnewline
\cmidrule{2-5} 
 & \emph{A} & B & C & D\tabularnewline
\midrule
\everypar{\hangindent1em \hangafter1}\raggedright Sponsor a conference
to explore the status of LGBT people in physics. & 1 & 3 & 5 & 2\tabularnewline
\addlinespace
\everypar{\hangindent1em \hangafter1}\raggedright Set up a committee
along the lines of the CSWP (Committee of the status of women in physics)
to represent LGBT people in physics. & 7 & 8 & 8 & 9\tabularnewline
\addlinespace
\everypar{\hangindent1em \hangafter1}\raggedright Conduct or facilitate
a national survey to gauge the needs of and present climate for LGBT
people in physics. & 16 & 12 & 10 & 14\tabularnewline
\addlinespace
\everypar{\hangindent1em \hangafter1}\raggedright Set up a subcommittee
within the broader Committee on Minorities. & 3 & 5 & 6 & 4\tabularnewline
\addlinespace
\everypar{\hangindent1em \hangafter1}\raggedright Develop \textquotedbl{}best
practice\textquotedbl{} guidelines for Department Chairs to create
an LGBT-inclusive environment. & 10 & 10 & 9 & 11\tabularnewline
\addlinespace
\everypar{\hangindent1em \hangafter1}\raggedright Something else
(with space for ideas) & 4 & 2 & 1 & 3\tabularnewline
\bottomrule
\addlinespace
\end{tabular}
\end{table}

\end{document}